\documentstyle [12pt,epsfig]{article}
\begin{document}
\parindent=0cm
\def\z{\u z}
\def\s{\u s}
\def\S{\u S}
\def\w{$\sim$\,}
\def\ic{\^\i }
\def\cc{\c{c}}
\def\ii{\"\i}
\def\arr{{$\Rightarrow$}}
\def\as{ast\'ero\"\i de}
\def\frac{$''$\hspace*{-.1cm}}
\def\ab{$\sim$}
\def\deg{$^{\circ}$\hspace*{-.1cm}}
\def\min{$'$\hspace*{-.1cm}}
\def\x{$\times$}

{\Large \bf A concise review of the Iranian calendar} \\

\vspace*{.5cm}

M. Heydari-Malayeri \\

Paris Observatory, \\
{\footnotesize 61 Avenue de l'Observatoire,\\
75014 Paris,\\
France.\\
m.heydari@obspnm.fr}\\

\vspace*{1cm}

%\fbox{Preliminary version, \underline {not} for release!}\\

\vspace*{1cm}

{\bf 1. Introduction}\\

\parindent=1cm 

The Iranian, or Persian, calendar is solar, with the particularity 
that the year is defined by two successive, apparent passages of the Sun 
through the vernal (spring) equinox. It is 
based on precise astronomical observations, and moreover uses a sophisticated 
intercalation system, which makes it more accurate than its younger  
European counterpart, the Gregorian calendar. 
It is currently used in Iran as the official calendar of the country. 
The Iranian calendar has a long, 
multi-millennial history with deep roots in the Persian culture, 
in particular the Zoroastrian heritage. 
However, this paper does not intend dealing with the history of the 
Iranian time reckoning since its most remote origins, because the 
questions addressed do not make it necessary. Interested readers can find 
detailed information on this  subject in,  
e.g., Taqiz\^adeh (1938), 
Birashk (1993), and Encyclopedia Iranica. 
 The present version of the calendar, used 
mainly in Iran, results from a reform led by the famous astronomer, 
mathematician, and poet Omar Khayy\^am (1048-1131 Christian era, 
hereafter A.D.: {\it Anno Domini}) which took place at 
the vernal equinox of A.D. 1079 (Friday, 21 March), corresponding to 
the 1st of Farvardin of A.P. ({\it Anno Persico}) 458.  
The  calendar was also called {\it Jal\^ali}, from the name 
of the ruler who ordered the reform. In modern times, 
its principles were officially adopted by the Iranian parliament on 31 
March 1925 (11 Farvardin  A.P. 1304). \\

The new year festival, called {\it Nowruz}, from Persian 
{\it now} ``new'' + {\it ruz} ``day'', is also a Zoroastrian legacy with 
many cultural significances, symbolizing the awakening or  
rebirth of the natural life after barren winter (see below Note 1). 
It is accompanied, several days or weeks before and 
after the spring  equinox, by  joyful ceremonies carrying centuries-old 
traditions.
 We can, for example, mention the house cleaning and seed sprouting 
before the Nowruz arrival and the fire festivity on the eve of the last 
Wednesday of the year as well as the mass picnic to verdant countryside 
on the 13th day of the new year. There are other celebrations, but their 
discussion  is out 
of the scope of this paper. Nowruz has been celebrated for several 
millennia by all the peoples  living in the Western and Central Asia 
(Afghans, Azeris, Caucasians, Kazakhs, Kurds, Kyrgyz', Tajiks, 
Turkmens, and others) regardless of ethnicity, 
religion, or language. Apart from its profound message, the 
non-ethnic and non-religious characteristics of  Nowruz seem 
to be the main reasons why it has been cherished as a common cultural 
heritage  by so many peoples with different backgrounds. \\

In spite of its outstanding status, mainly its precision and
rationalistic characteristic of relying on detailed astronomical
observations, the Iranian calendar is poorly known in the West. In
fact, few studies have so far analyzed the basic elements of its
system in the light of modern astronomical findings, although several
works have addressed the historical origins of the calendar. 
In particular, the true value of the length of
the year in the Iranian calendar should be emphasized, since nowadays
there is a widespread confusion between the concepts of the modern
``tropical'' year and the vernal-equinox year upon which the Iranian
calendar is based. A misinterpretation of this topic may upset the
intercalation system and affect the accuracy of the calendar. One of the
goals of this paper is therefore to clarify this point. \\

The paper is not only aimed at calendar experts but effort has been
made to be accessible by non specialists. It is organized as
follows. After this Introduction, Section 2 presents a general
description of the Iranian calendar. Then, in Section 3, we define the
year and discuss its length using the results from recent
astronomical research. In Section 4 we underline the difference
between the ``tropical'' and Iranian years, while the intercalation
system, based on the 33-year cycle, is described in Section
5. Although the paper is mainly interested in the astronomical
background of the Iranian calendar, historical aspects are not
overlooked and, in particular, Section 6 presents a brief note
regarding the history of the reform led by Khayy\^am. In recent years,
some Iranian calendar experts have suggested the presence of a
2820-year cycle in the Iranian calendar. We discuss this issue in
Section 7 and point to the flaws of this scheme.
The correspondence between the Iranian and other calendar
systems is dealt with in Section 8. And, finally, the concluding 
remarks are given in Section 9.  Throughout the paper,
complementary explanations are given in 16 notes which appear at
the end of the article.\\

\vspace*{1cm}

\parindent=0cm

{\bf 2. General description} \\

\vspace*{0cm}

\parindent=1cm

Nowruz starts at the precise instant when the Sun, in its
apparent annual course on the sky, coincides with the vernal 
equinox, an event that can occur at any time during the 24-hour 
diurnal period. The vernal and autumnal equinoxes are defined as 
the points of intersection between the ecliptic (the apparent path of 
the Sun on the sky) and the celestial equator (the projection of 
the equator of the Earth on the sky), although the  
vernal equinox is the reference point from which the right ascensions 
(equator system) and the longitudes (ecliptic system) of the 
heavenly bodies are measured (Note 2). The vernal equinox is also 
the moment when the Sun appears to cross the celestial equator 
heading northward. However, nowadays it is more conveniently defined 
as the instant when the Sun's ecliptic longitude is zero degrees.
The Nowruz event is at present measured to an accuracy of better 
than 1 millisecond (Malakpour 2004).  
Using the ephemeris calculated by the French  
{\it Institut de M\'ecanique C\'eleste et de Calcul des 
Eph\'em\'erides} (IMCCE)   for the Gregorian 
interval A.D. 1583-2500, we find that 584 equinoxes, or about 64\% of the 
whole events, occur on March 20 for the Tehran longitude. 
The distribution among the 
neighboring dates is presented in Table 1. Each year the vernal equinox 
occurs later, with a delay usually under 6 h, with respect to the 
preceding one. The delay is equal to the fraction  of the day which 
exceeds the year of 365 entire days. The date  of the vernal equinox 
does not follow this systematic forward shift because the accumulated 
delays are absorbed in the Iranian calendar by the leap day added to 
the calendar every four or five years (see below, Section 5). \\

\vspace*{0cm}

The IMCCE data allowed us also to calculate the shift  
{\it dt = t(n+1) - t(n)} for the interval  +1000 to +2500, $t$ 
being the instant of the vernal equinox for year $n$. 
This resulted in the average value of  5.81662 h  $\pm$ 0.00240 
(standard deviation) or 5h 48m 59.83s $\pm$ 8s (s.d.), for the 
interval A.D. 1000-2500. The lower and upper bounds, 5.54111 and  
6.06444 h belong to years 1095 (A.P. 474) and  1216 (A.P. 595) 
respectively. The plot showing the variation of the delay, {\it dt}, 
over the years is displayed in Figure 1. Since the length of the 
vernal-equinox year is not constant (see below, Section 3), the average shift 
depends on the interval considered. Table 2 shows the  shift for several 
different intervals, as well as the corresponding average year 
length referred to two successive vernal equinoxes, in real solar time.

\vspace*{.5cm}

\begin{table}[htb]
\begin{center}
\caption[]{{\bf  Iranian equinox dates for period A.D. 1583-2500  }}
\vspace*{0.5cm}
\begin{tabular}[h]{|c|c|c|}
\hline
March date & Event frequency & Percentage (\%)\\
\hline
18 &    0 &     0 \\
19 &    33 &    3.6 \\
20 &    584 &   63.6 \\
21 &    301 &   32.8 \\
22 &    0   &   0 \\
\hline
\end{tabular}
\end{center}
\end{table}

\vspace*{1cm}

\begin{figure}
\begin{center}
\epsfig{figure=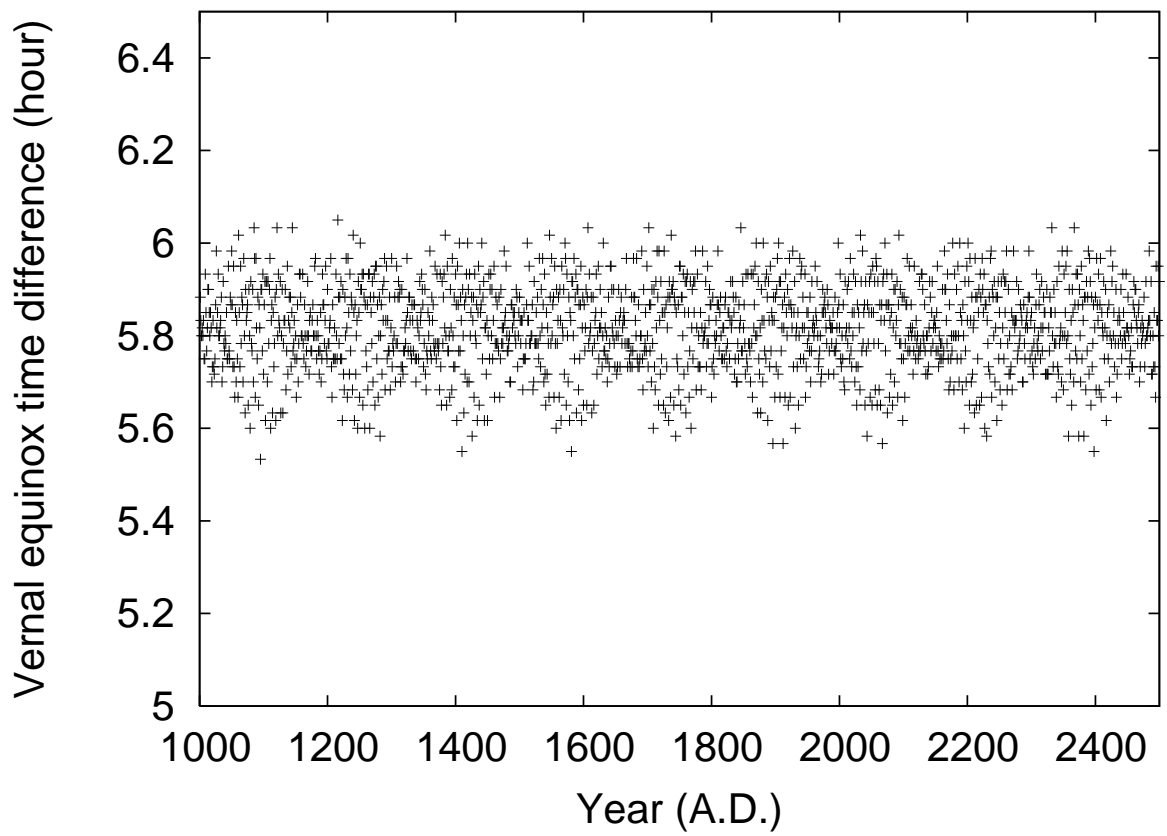, width=\linewidth}
\caption{{\it The variation of  the time delay between  
a vernal equinox event with respect to the preceding one for the 
period A.D. 1000-2500.  Mean value:  5h 48m 59.8s.  } }
\end{center}
\end{figure}

Obviously, the first day of the calendar year cannot
start at the precise time of the vernal equinox and should begin at
midnight  (Tehran true time), the beginning of the day according to 
 a long Iranian tradition (Note 3).  
Consequently, if the vernal equinox falls before noon on a
particular day, then that day is the first day of
the year. If the vernal equinox occurs after noon, the following day 
begins the calendar year. In other words, the year 
begins at the midnight closest to the instant of equinox. In the 
past this was achieved by observing noon altitudes of the Sun 
and declaring  Nowruz to be the day when the solar altitude first 
exceeded the altitude of the celestial equator. 
Although the calendar year can start with a delay or advance of 
less than 12 h with respect to the equinox event, the exact instant of 
the vernal equinox, or the year transition ({\it tahvil-e s\^al}), 
has utmost cultural importance. No matter 
the moment during day or night, all family members,   
cleansed and dressed in new garments, wait the astronomical occurrence  
around a  ceremonial table ({\it haft-sin}) decorated 
with several symbolic items. Nowadays the instant of Nowruz is aired by TV 
and radio stations, whereas in the past canon shots or wind 
instruments and drums were used to mark the event. When the exact 
time arrives, there is a burst of rejoicing and cheering and greetings 
are exchanged. \\

\begin{table}[htb]
\begin{center}
\caption[]{{\bf Average time delay between two successive vernal equinoxes}}
\vspace*{0.5cm}
\begin{tabular}[h]{|l|l|l|}
\hline
Epoch interval  & Shift        & Average year \\
                & {\it (hour)}       & {\it (solar days)} \\
                & {\it (h, m, s)}    &             \\
\hline
--4000 to +2500  & 5.81458      &  365.242274 \\   
                & 5h 48m 52.5s &             \\
0 to +2500      & 5.81603      &  365.242335 \\
                & 5h 48m 57.7s &             \\
+1000 to +2500  & 5.81662      &  365.242359 \\ 
                & 5h 48m 59.8s &             \\
+1500 to +2500  & 5.81669      &  365.242362 \\         
                & 5h 49m 0.1s  &             \\
+1800 to +2200  & 5.81537      &  365.242307 \\   
                & 5h 48m 55.3s &             \\
+800 to +1200   & 5.81609      &  365.242337 \\           
                & 5h 48m 57.9s &             \\
+1995 to +2005  & 5.83087      &  365.242953 \\    
                & 5h 49m 51.1s &             \\    
\hline
\end{tabular}
\end{center}
\end{table}

%\vspace*{.5cm}

The starting point of the current Iranian
calendar is the vernal equinox occurred on Friday March 22 of the 
year A.D.  622. Historically, on 22 September of that year (7th of the 
Arabic month Rabi' I), the Islam's prophet emigrated to Medina 
({\it Hijra}). However, the second Caliph Umar ibn al-Khatab, 
when adopting a calendar system around A.D. 638, preferred to take 
the starting point of the Islamic lunar calendar two months and eight 
days back, at the beginning of the Arab sacred month of Moharram, 
corresponding to July 19, A.D. 622. On the other hand, 
the fact that in the lunar calendar the months are not 
related to the natural seasons and  the dates move with respect to the solar 
cycle makes the lunar calendar inadequate for the civil administration 
(for example for agriculture scheduling and harvest-tax collection). 
Therefore,  the troubles brought by the lunar calendar were 
strongly felt in Iran four centuries later during the reign of 
Jal\^al ad-Din Malek Shah, of the the Saljuqid dynasty. Moreover, 
the Nowruz celebration had slipped into the middle of the Pisces sign, 
due to neglecting the intercalations (see below, Section 5), 
and in fact, according to early historians and astronomers, the 
main purpose of the reform was to fix Nowruz at the vernal equinox 
(Encyclopedia Iranica). \\

For the reasons mentioned above, the brilliant vizier and  political figure 
Nez\^am-ol-Molk persuaded Malek Shah to reform the calendar. He appointed  
a  group of astronomers headed by Omar Khay\^am (Note 4) who organized  
an astronomical observatory in the capital city Isfahan 
(some sources have mentioned Rey or Neysh\^apur). On the basis 
of observations and calculations, the reform panel adopted the vernal 
equinox of the year A.D. 622, which places the beginning of  
the Iranian calendar six months before the {\it Hijra}. Similarly,  
it  invented  a unique intercalation system based on a 33-year  
cycle, which will be described below (Section 5).  
Furthermore, the reform panel adopted 12 equal months of 30 days 
each, and inserted the remaining 5 or 6 ``stolen days'' or 
{\it andarg\^ah} (epagomena) between 30 Spandarmad (Esfand) and 
Nowruz, as was practiced in the Iranian pre-Islamic calendar.  
And, interestingly, the panel revived the ancient Persian month names 
(Table 3), although other alternatives were considered and new Persian 
poetic names were proposed. \\

\vspace*{.5cm}

\begin{table}[htb]
\caption[]{{\bf Iranian calendar months and seasons}}
\vspace*{0.5cm}
\begin{tabular}[h]{|c|c|c|c|c|c|}
\hline
Order & Avestan & Middle Persian & Modern Persian & Days & Seasons \\
      & (A.D. {\it c.} --2000 to --300) & (A.D. {\it c.} --300 to +700) & & & \\
      &     & & & & \\
\hline
1  & Fravashi/Fravarti &  Fraward\ic n &  Farvardin & 31 & Spring\\
   & {\it (Devine essence)}  &            &            &    & \\
\hline      
2  & Asha Vahishta     &  Ardawahisht & Ordibehesht & 31 & Spring\\
   & {\it (Best righteousness)} &       &             &         &    \\
\hline 
3  & Haurvat\^at & Khord\^ad & Khord\^ad & 31 & Spring \\
   & {\it (Wholeness, integrity)} &       &           &     &          \\
\hline 
4  & Tishtrya          & T\ic r        & Tir & 31  & Summer       \\
   & {\it (Sirius, rain star)} &           &          &         &           \\
\hline 
5  & Am\'er\'et\^a & Amurd\^ad & Mord\^ad/  &     31 & Summer \\
   & {\it (Immortality)} & & Amord\^ad & & \\
\hline 
6  & Khshathra Vairya & Shahrewar & Shahrivar & 31 & Summer \\
   & {\it (The good dominion}  & & & & \\
   & {\it of choice)}               & & & & \\
\hline 
7  & Mithra & Mihr & Mehr & 30 & Autumn \\
   & {\it (Sun, friendship,}  & & & & \\
   & {\it promise)}             & & & & \\
\hline 
8  & Ap & \^Ab\^an & \^Ab\^an & 30 & Autumn \\
   & {\it (Water)} & & & & \\
\hline 
9  & \^Athra & \^Adur & \^Azar & 30 & Autumn \\
   & {\it (Fire)} & & & & \\
\hline 
10 & Dathush\^{o} & Day & Dey & 30 & Winter \\
   & {\it (Creator)} & & & & \\
\hline 
11 & Vohu Manah & Wahman & Bahman & 30 & Winter \\
   & {\it (Good Mind)} & & & & \\
\hline 
12 & Sp\'ent\^a \^Armaiti & Spandarmad & Esfand & 29/30 & Winter \\
   & {\it (Holy serenity)} & & & & \\
\hline
\end{tabular}
%\end{center}
\end{table}

\newpage
 
%\vspace*{.8cm}

We notice therefore that the reform's objective was not only resolving  
administrative and economical problems,  but also aimed at 
preserving Nowruz, one of the main symbols of the Iranian identity. 
In brief, the reform had an outstanding consequence, probably 
unprecedented in the Iranian history. It attached Nowruz tightly 
to the vernal equinox, 
which  was not the  case in the Sasanian Zoroastrian calendar, 
because the latter did not apparently employ the 4-year intercalations, 
but instead added a supplementary month to the calendar every 120 
years (de Blois 1996).  \\

In contrast to the Islamic calendar, which is based on the 
lunar synodic month and the corresponding lunar year of 
of 354 days, the Iranian year is divided into 12 months 
with tight links to the real, annual seasons.  
The first 6 months of the modern calendar have 31 days, 
the next 5 months 30 days and the last one 29 or 30 days. 
This scheme of month lengths is in accord with the fact that the 
astronomical seasons do not have an equal number of days, 
spring and summer being longer than autumn and winter (Note 5).  
The month names, in their original Avestan form (Note 6), their 
meanings, and  their linguistic evolution over the ages are presented in  
Table 3.\\

\vspace*{1cm}
\parindent=0cm

{\bf 3. The length of the year }\\

The Iranian year is a  ``tropical year'' with the vernal equinox as 
its  reference point. However, it should not be confounded with the
tropical year set up by modern astronomers (Note 7). Nowadays, 
the tropical year is defined  as 
the interval during which the Sun's {\it mean} longitude, 
referred to the {\it mean} equinox of the date, increases by 
360 degrees. This definition  was 
adopted by the International Astronomical Union at its General 
Assembly in Dublin in September 1955 (Seidelmann et al. 1992).  
Several researchers, notably Meeus \& Savoie (1992), Cassidy (1996), 
and Meeus (2002) have underlined the difference between the 
vernal-equinox year and the newly introduced tropical year. 
Nonetheless, the mistake of equating the two is unfortunately 
largely propagated.  
In fact most astronomers and calendar adepts are defining the tropical 
year as starting with the vernal equinox but using the length of 
the newly-defined tropical 
year  for their calculations (Note 8). And unfortunately some Iranian 
astronomers and calendar experts are not exempt from this confusion. \\

\parindent=1cm

It should be emphasized that this new concept of the tropical year, 
introduced for celestial mechanical studies, is based on the mean 
longitude of the Sun and does not depend on a specific origin for 
the annual apparent motion of the Sun. 
It aims at  the long-term behavior of the year by 
taking into account the precession (secular terms) and not 
short-term, periodic gravitational disturbances. 
No matter how useful  this concept, the  
traditional solar calendars, Iranian as well as Gregorian (Note 9), 
are based on the mean length of  the real vernal-equinox year. 
The situation of the nomenclature as it is now lends itself  
to confusion, because the same term (tropical year) is used 
for two different, but very closely related, concepts. Therefore, it would be 
highly expedient for the International Astronomical Union to 
remove the ambiguity by adopting two distinct terms for both notions. 
It would be quite logical to apply ``tropical year'' only in its 
modern, celestial mechanical meaning, and use the term ``vernal-equinox 
year'' for the interval between two successive passages of  the 
Sun through the vernal equinox. Naturally, Iranians call it 
the {\it Nowruz year}.   \\

\parindent=1cm

Let us look more closely into the difference between the tropical and 
vernal-equinox years. Bretagnon and Rocher (2001) give the following 
expression for the length of the mean tropical year:

\begin{center}

{\it  365.24219052 -- 61.56 10$^{-6}$ T -- 68.4 10$^{-9}$ T$^{2}$ 
   + 263.0 10$^{-9}$ T$^{3}$  + 3.2 10$^{-9}$ T$^{4}$ }
   \hspace*{1cm} [1] \\
\end{center}

\parindent=0cm

where $T$ is the barycentric  dynamical time (TDB) or, more simply,   
uniform (ephemeris) days of 86400 seconds (International Atomic Time, 
TAI), counted in Julian millennia (of 365250 days) from the present epoch. 
The expression is valid over an interval of  $\pm$10,000 years 
from J2000.0. Expressed in universal time, based on the Earth's 
rotation about its axis, the same authors give the 
following polynomial formula for the tropical year:

\begin{center}
{\it 365.2421789 -- 135.63 10$^{-6}$  T -- 68.4 10$^{-9}$ T$^{2}$ 
   + 263.0  10$^{-9}$ T$^{3}$ + 3.2 10$^{-9}$ T$^{4}$ }
   \hspace*{1cm} [2] \\
\end{center}

where $T$ here  is  universal time.  We see therefore 
that at the epoch of J2000 the length of  the tropical year was
365.24219052  uniform days of 86400 seconds, which is equal to 
365.2421789 real solar days, and can be rounded off to 365.2422 days. 
The formulas [1] and [2] also indicate that the length of the tropical 
year is not constant. As shown also in Figure 2, it is decreasing by about 5  
milliseconds per year, because the precession is presently 
speeding up. In addition, not shown in Figure 2, its length in real 
solar days decreases by about another 6 millisecond 
per year as a result of the slowing down of the  
Earth's rotation about its own axis,  mainly due to the 
tidal effect of  the Moon.  It should be underlined that 
presently we are unable to accurately predict the length of 
the day. The reason is that the Earth's rotation 
undergoes unpredictable, irregular variations over different 
time scales, as a result of both internal and external geophysical 
processes and astronomical perturbations.  \\

\hspace*{1cm}
On the other hand, the mean time interval between two successive 
vernal equinoxes, as derived by Meeus (2002) for the true longitude 
of the Sun, based on the orbital elements of the Earth 
(Simon et al. 1994) and the exact solution of Kepler's equation,  
is represented by the polynomial expression:

\begin{center}

{\it 365.2423748 + 10.34 10$^{-5}$ T -- 12.43 10$^{-6}$ T$^{2}$ 
   -- 22.63  10$^{-7}$ T$^{3}$ + 1.31 10$^{-7}$ T$^{4}$ }
   \hspace*{1cm} [3] \\

\end{center}

for the period ranging from about 500 B.C. to A.D. 4500, where
$T$ is measured in Julian millennia of 365250 (ephemeris) uniform 
days from the epoch J2000.0. It should be underlined that this 
expression represents the global evolution of the real  
vernal-equinox year over large time-spans by allowing for  
the non-uniform Keplerian motion of the 
Earth.  It does not, however, take into account the 
fluctuations in the length of the year caused by various periodic 
gravitational perturbations. The exact solution is obtained in the 
IMCCE ephemeris by an equation composed of 96 terms.  
The above equation [3] can be expressed in solar days, as:

\begin{center}

{\it 365.2423632 + 2.93 10$^{-5}$ T -- 12.43 10$^{-6}$ T$^{2}$ 
   -- 22.63 10$^{-7}$ T$^{3}$ + 1.31 10$^{-7}$ T$^{4}$ } 
   \hspace*{1cm}  [4] \\
\end{center}

\parindent=1cm

The comparison between equations [3] and [1] or [4] and [2] shows clearly 
that the mean vernal-equinox and tropical years do not have the same 
length. For example, for the epoch of +2000, the mean tropical year is 
365.2421789 days, whereas the mean vernal-equinox year is 365.242362 
days, that is 15.82 s longer. In the same way that the tropical year 
was rounded off to 365.2422 days, the vernal-equinox year can be 
approximated to 365.2424 days. Moreover, it is obvious from 
equation [3] that the vernal-equinox year also varies over the ages 
(see also Figure 2). Around the epoch --3000 it was 365.241872 uniform days, 
whereas and at the beginning of the Christian era it was increasing 
to 365.242138 uniform days. It is still increasing and will attain its 
largest value, 365.242525 uniform days, in the epoch +5000.
Therefore, the difference between the vernal-equinox and tropical 
years is not limited only to their lengths, since their long-term 
variations over the time is also dissimilar. In contrast to the 
vernal-equinox year, the tropical year will continually decrease 
at least until the epoch +8000.  \\

Figure 2 also displays the variation of the real 
vernal-equinox year, 
in solar days, obtained by the IMCCE calculations. Those data are shown in 
two different forms for the sake of better visibility, since the year exhibits 
considerable fluctuations over relatively short periods.  
The discrete points, centered on vertical bars, represent 
a mean value for a 500-year interval with the bars showing the 
standard deviation over that interval. 
The smallest range of the fluctuations  belongs to the epoch 0 where the 
vernal-equinox varies over about 20s, while for the epoch +2000 the 
spread is twice larger. 
The undulating curve displays the IMCCE data after being smoothed 
out by a filtering method in which all the variations with 
periods under 100 years are removed. We see therefore that the 
main fluctuations in the length of the vernal-equinox year result 
from relatively short-period perturbations. The overall shape of the 
real vernal-equinox year nears that of the mean vernal-equinox 
year if we use a stronger smoothing technique. 
Note that in order to convert uniform days into  solar 
days one must use a correction parameter, called the derivative of 
Delta-T, which 
can be represented by the difference between equations [1] and [2].  
Generally, the correction increases the uniform-day years before the 
epoch +2000.0 and reduces those after that epoch (Note 10). \\

\begin{figure}
\begin{center}
\epsfig{figure=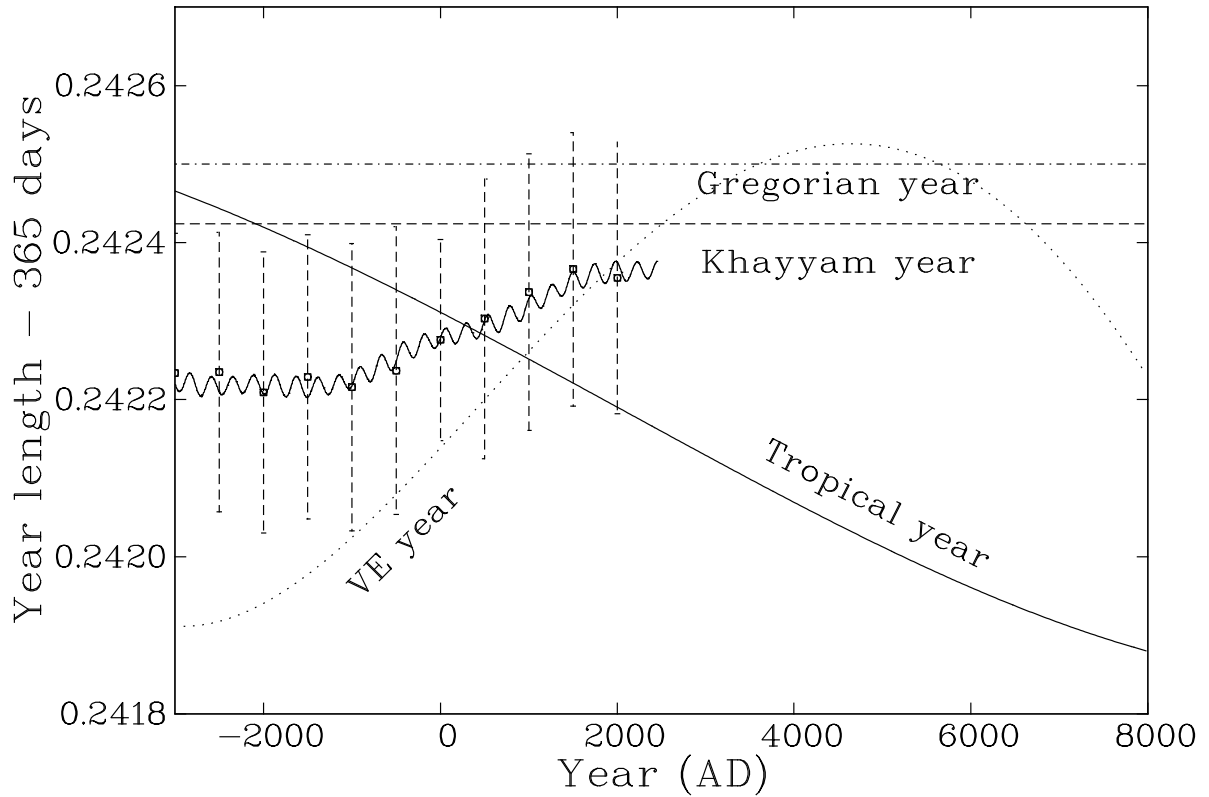, width=\linewidth}
\caption{{\it The lengths of the tropical and  vernal-equinox years   
varying over the ages, shown in full and dotted curves
respectively, both in units of uniform (ephemeris) 
days. The duration of the real vernal-equinox year, in solar days, 
or the Iranian year,  
obtained by the IMCCE ephemeris, is displayed in two manners.
The discrete points  represent mean 
values over 500-year intervals with their corresponding standard 
deviations shown by the vertical bars. Similarly, the wavy  
curve displays a smoothed version of the ephemeris data in 
which fluctuations with periods smaller than  100 years are 
withdrawn. The horizontal lines show the lengths of the Khayy\^am and 
Gregorian calendar years. They are meant for comparison with the 
vernal-equinox year lengths in solar days represented by 
the IMCCE data. }}
\end{center}
\end{figure}

As it was pointed out above, the length of the real vernal-equinox 
year can deviate considerably from the mean value over short 
periods mainly due to the gravitational pull of the Moon and the  
planets.  
The fluctuations of the year length are also visible in  Figure 1. 
Over the interval A.D. 1000-2500 the shortest year,  
365.230880 days, is A.D. 1095 and the longest A.D. 1216 with 
365.252685 days, or 365d 6h 3m 52.0s. The mean length depends on 
the interval considered, as displayed in Table 2. We see that 
around Khayy\^am's epoch the mean duration of the 
vernal-equinox year over the 400 year interval A.D. 800-1200
was 365.242337 solar days, which is also the average value 
for the whole period 0 to +2500.  In the following we will 
use a mean value of  365.242362 days 
(corresponding to the period +1500 to +2500) for the 
length of the vernal-equinox year in the coming centuries.   \\

\newpage

%\vspace*{1cm}

\parindent=0cm

{\bf 4. The difference between the ``tropical'' and Iranian years } \\

\parindent=1cm
Here we give more explanations in order to highlight the fact that 
the Iranian vernal-equinox year is not equal to the newly introduced 
``tropical year''. Since the Earth's orbit around 
the Sun is elliptical, its orbital velocity is not uniform. At the 
perihelion of its orbit the Earth is the closest to the Sun, and 
therefore  moves faster than average, while at aphelion, when it is 
the farthest away from the Sun,  it moves slower. Moreover, due to 
the precession, the vernal equinox slowly regresses along the 
ecliptic by 50.3 arcseconds per year  with respect to stars, while 
the major axis of the Earth's orbit rotates  in the direct sense by 
11.6 arcseconds per year. Consequently, when the Sun crosses the 
vernal-equinox point, spending a year in its apparent path along the 
ecliptic,  the Earth has not made a complete circuit. 
It has made a complete journey on a distorted orbit 
(due to the rotation of the orbit) less a small arc.  This small arc
is covered with a somewhat different speed, according to its position
with respect to the perihelion of the Earth's orbit. As a result, 
depending on the starting point adopted for the ``year'', one complete 
circuit in tropical longitude will have a somewhat variable length. 
This means that the length of a ``real'' tropical year depends upon the 
chosen reference point on the ecliptic. \\

A simple mathematical formula can facilitate the explanation. 
The year length can be represented by: 
{\it Y = T - t}, where $T$ is the time needed by the Sun for making a 
complete circuit with respect to the perihelion 
and $t$ the time gained because the reference 
point approaches the Sun. The  time gained is represented by 
the ratio of the arc length, 
$S$, and the apparent velocity of the Sun: {\it t = S/V}.
But since the arc length is constant, $t$ depends only on the 
velocity. For example, at the December solstice the Earth 
is near the perihelion (to which it arrives in early January) 
and the Sun apparently moves faster than 
average towards the solstitial point. Hence the time gained  
is smaller than that for the vernal equinox, and 
the tropical year as measured for this point will not only be longer than 
that for the vernal equinox, but the longest of the four cardinal years.
On the basis of the planetary ephemeris calculated by Simon et al. (1994), 
Meeus (2002) has derived the mean intervals between two successive 
passages of the Sun through the four seasonal points for several epochs. 
The results, listed in Table 4, clearly show that these various ``tropical'' 
years do not have the same length (Note 11). And, more particularly, 
the vernal-equinox year is different from the ``tropical'' year. 
 In fact the tropical year represents  
the mean value of the four seasonal figures (Note 12). 
It should be emphasized that  the year lengths 
listed in Table 4 are in uniform days, 
and must be converted into solar days for calendar use, as 
listed in the second row for the vernal-equinox year (Note 10).   \\

\begin{table}[htb]
%\begin{center}
\caption[]{{\bf Various ``tropical'' years for several epochs\,$^\ast$}}
\vspace*{0.5cm}
\begin{tabular}[h]{|c|c|c|c|c|}
\hline
Epoch & Vernal-equinox & Summer-solstice & Autumnal-equinox &  
     Winter-solstice \\
      & {\it Nowruz} year          &       year &       year & year      \\
\hline
--1000 & 365.242022 & 365.241859 & 365.242718 & 365.242871\\
      & 365.24223  &            &            &            \\
\hline
0     & 365.242138 & 365.241726 & 365.242496 & 365.242883\\
      & 365.24227  &            &            &           \\
\hline
+1000 & 365.242261 & 365.241648 & 365.242257 & 365.242840\\
      & 365.24232  &            &            &           \\
\hline
+2000 & 365.242375 & 365.241627 & 365.242018 & 365.242741\\
      & 365.24236  &            &            &           \\
\hline
+3000 & 365.242464 & 365.241659 & 365.241800 & 365.242594\\
      & 365.24237  &            &            &           \\ 
\hline
+4000 & 365.242516 & 365.241734 & 365.241620 & 365.242407\\
      & 365.24235  &            &            &           \\
\hline
\end{tabular}
%\end{center}
%\end{table}
\begin{center}
{\small\,$^\ast$ The year lengths are in ephemeris (uniform) days, except for 
the vernal-equinox year which is given also in real solar days.}
\end{center}
\end{table}

The length of the year in the Iranian calendar, as conceived by  
Khayy\^am et al., is 365.2424.. days (Youschkevitch \& Rosenfeld 1973), 
a logical consequence of the intercalation system: 
365 + 8 / 33 = 365.2424.. (see Section 5).  
This is in  good agreement with the length of the year at 
his time,  365.2423 days. Khayy\^am's estimate 
of the year length  agrees even better  with the present-day 
mean value of 365.2424 days. Compared to the Gregorian year of 
365 + 1 / 4 -- 1 / 100 + 1 / 400 = 365 + 97 / 400 = 365.2425 days, 
Khayy\^am's value was based on a more accurate knowledge of the solar 
annual motion. Note that recently an extremely precise value 
for the duration of the year 
(365.24219858156 days) has been attributed to Khayy\^am 
%(<A HREF="http://www-gap.dcs.st-and.ac.uk/~history/Mathematicians/Khayyam.html">
(O'Connor \& Robertson 1999), although it is supported by   
no historical source, as far as we know. We think that this may be a 
spurious value resulting from the erroneous 2820-year cycle suggested 
recently (see below Section 7). \\

The precise knowledge of the year duration is one thing, using it 
practically to construct a calendar system is another, because 
the calendar year of 365 entire days is not  equal to  
the real solar year. A good calendar is therefore the one 
whose intercalation system yields a year length the nearest 
possible to the reality.  \\

\parindent=0cm 
\vspace*{.5cm}

{\bf 5. The intercalation system}  \\

\parindent=1cm 

The average time shift, {\it dt}, between two successive vernal 
equinoxes, listed in Table 2, indicates also the excess time with 
respect to a calendar year of  365 full days. In other words, each 
year the solar time leads by {\it dt/24} day with respect to 
the calendar. Taking {\it dt} = 5.81662 h for the interval
+1000 to +2500,  
this lead  will be $5.81662/24 = 0.242359 \simeq 0.2424$ days per year. 
Thus, after four years, the advance amounts to 4 \x\, 0.24236 = 0.96944 
days, which is approximated to 1 day, necessitating a leap year of 366 
days. This correction of 1 day every 4 years is though too generous 
and therefore  results in a lead of the calendar year 
over  the solar year. This point constitutes the fundamental problem 
of the solar calendars, and this is the reason why in the Iranian 
calendar the intercalation is not carried out systematically every 4 
years (see below). On the other hand, if no intercalations are applied, 
the Nowruz feast will regress with respect to the 
vernal equinox, moving through all the seasons in about 1506 years. 
More explicitly, after four years the vernal equinox will 
happen on Farvardin 2 and after eight years on Farvardin 3, and so on.
At the same time, Nowruz will slip more and more into  winter. \\

The Iranian calendar has a basic, 33-year intercalation cycle, 
which consists of 33 \x\, 0.2424 = 7.9992 = 8 leap years of 366 days and 
25 ordinary years of 365 days. There are two kinds of leap years: 
i) leap year 
after 3 ordinary years (called {\it quadrennial}), ii) leap year 
after 4 ordinary years ({\it quinquennial}). During a quadrennial  
period, the vernal equinox occurs as two pairs systematically 
two times after noon  then two times before noon. However, about  
every 33 years a vernal equinox  happens  very close to 
midnight, coming between two such pairs and forming a quinquennial 
period. More explicitly, this situation happens dominantly  every 33 years 
and sometimes, much less often, after an interval of 29 years. \\

The first leap year of the 33-year cycle is quinquennial  and the 
seven others quadrennial. The intercalations are therefore  applied 
to the years: 5, 9, 13, 17, 21, 25, 29, 33.  The total cycle
amounts to 25 \x\, 365 + 8 \x\, 366 = 12053 calendar days.  In comparison,
the solar time duration is currently 33 \x\, 365.242362 = 12052.99795 days.  
This
means that  a 33-year cycle  advances by 0.00205 days, or about 3  
minutes, with respect to the solar time. This is the reason why a 
strict 33-year cycle needs future adjustments. Assuming a constant 
length for the vernal-equinox year, after about 500 cycles, or about 
16,000 years, the advance will add up to 1 day.  
However, since the duration of the year  decreases after 
the epoch +3000 and would reach a value of about 365.24182 days in 
the epoch +8000, a shorter period is required in order that the 
advance attain one full day. Anyhow, when the error totals 1/33 of 
a day, a 33-year cycle needs to be shortened to 29 years in order 
to keep the calendar in pace with the solar time.
It should be emphasized that detailed 
extrapolations over 
very long periods are unavailing because the year length is not 
constant and moreover unknown perturbations may add up in the 
distant future to have a considerable effect.  \\

In the past, when  Newton had not yet put forward his theory  
of gravitation and Laplace, Lagrange, Euler, Hamilton, and others 
had not contributed to the formulation of  celestial mechanics,  
the precise instant of the vernal equinox was predicted by extrapolating 
detailed observations. Today, thanks to 
the remarkable improvements in the models  of orbiting bodies, and more 
particularly the development of powerful computers and numerical 
methods, the instant of the vernal equinox can be predicted with 
very high precision.   
However, observational measurements are necessary  to check the actual 
correspondence between the mathematical model and the reality. 
The observations are also an important part of the Iranian calendar 
culture, and it would be wise to be preserved. Moreover,  now that 
amateur astronomy has made big technical advances, more 
people can participate in the observation of the solar motion  
around Nowruz. Since Nowruz is accompanied by several public 
ceremonies and festivities, the widespread observation of 
the Sun around the spring equinox can be a delightful modern addition. 
This will also provide a good opportunity for teaching  astronomy and 
calendar science to the public. The same thing can be recommended also for 
the observation of the solar motion during the {\it Mehreg\^an} 
festival at the beginning of Mehr, the 7th month, which celebrates  
the autumnal equinox (Notes 13 \& 14).    \\

\newpage
%\vspace*{1cm}
\parindent=0cm 

{\bf 6. A notice of historical interest} \\

\parindent=1cm 

The information about Khayy\^am's calendar reform come to us 
not directly from himself, but only through brief accounts by 
later astronomers. In fact 13 years after the instauration of the 
reform  Khayy\^am fell into disgrace  after Malek Shah's death and 
Nez\^am-ol-Molk's murder by  Assassins, a politico-religious order  
led by Hasan Sabb\^ah. Following 
the death of the king, his wife ruled as regent for two years, 
and Khayy\^am fell heir to some of the hostility she had demonstrated 
towards his patron, Nez\^am-ol-Molk, with whom she had quarreled   
over the question of royal succession. Moreover, 
orthodox Muslims, who disliked Khayy\^am because of his freethinking,  
evident in his quatrains, became highly influential at court. 
Maybe they were not happy neither with Khayy\^am's 
frank  moves for valuing the Persian identity.  
Financial support was cut from the observatory and its 
activities, among which the reform, came to a halt (Youschkevitch \& 
Rosenfeld 1973). \\

There is no doubt though that Khayy\^am and his group implemented the 
intercalation system based on quadrennial and quinquennial leap years. 
The astronomer Kh\^azeni (Note 4), who was probably 
a member of the reform group 
and later, during the reign of Sanjar, Malek Shah's son, 
authored the {\it Sanjari zij} (astronomical observations and tables), 
uses the quadrennial/quinquennial intercalations, which most probably 
resulted from the work done by Khayy\^am's reform group. For a period of 
220 years in the Jal\^ali calendar,  Kh\^azeni calculates  
53 intercalations,  8 quinquennial and 45 quadrennial (Note 15). 
About two centuries later, the eminent astronomer and mathematician 
%<A HREF="http://www-gap.dcs.st-and.ac.uk/~history/Mathematicians/Al-Tusi_Nasir.html">
Nasireddin Tusi (A.D. 1201-1274), director  of  Mar\^agha  
observatory who created the {\it Ilkhani zij}, and  is recognized  
for having made an exhaustive criticism of Ptolemic astronomy and having 
presented a new mathematical model of planetary motions, gives 
brief but important information about 
the Jal\^ali reform. In his {\it Ilkhani zij}, he clearly states that 
an intercalation is applied every 4 years, making a year of 366 
days, and after 7 or 8 quadrennial intercalations a quinquennial 
intercalation is performed on the basis of {\it induction}.
We see therefore an implicit reference to the scheme of 
29 or 33-year cycles. For information about the various 
intercalation attempts over  centuries,  based on the 
quadrennial/quinquennial design see Sayy\^ad (1981) who gives an 
interesting compilation on the subject. \\

\vspace*{1cm}
\parindent=0cm

{\bf 7. The putative 2820-year cycle } \\

\parindent=1cm

Recently some Iranian calendar experts, mainly the late  Behruz 
(1952) and the late  Birashk (1993), have suggested the existence 
of so far unknown  cycles in the Iranian calendar. In particular, they
divide the calendar into cycles of 2820 years, itself split in 21
subcycles of 128 years and 1 subcycle of 132 years (2820 = 21 \x\, 128 +
132).  The subcycles of 128 and 132 years are themselves divided into
periods of 29, 33, and 37 years (128 = 29 + 3 \x\, 33; 132 = 29 + 2 \x\, 33 + 
37). In brief, the cycle consists of 2137 ordinary years and 683 
leap years. \\

This scheme contains several drawbacks. It is claimed that the
2820-year cycle represents the interval after which the 
vernal equinoxes are repeated at the same instant of the day. The 
scrutiny  of the IMCCE ephemeris data does not allow us 
to uphold this claim, which has already been criticized  by calendar 
researchers, among them the Iranian astronomers Malakpour (2004) and 
Sayy\^ad (2000). Moreover, the whole 2820-year cycle gets a lead of 
about 0.5 days over the solar time. Using the present length of the 
vernal-equinox year, the cycle  amounts to 
2820 \x\, 365.242362 = 1029983.461 days, while the calendar will 
have 2137 \x\, 365 + 683 \x\, 366 = 1029983 days. In order that the  
cycle remain in step with the solar time, the year length must 
therefore be 365.2421985815603 days or 
about 365.2422 days. This figure, which  
results also from the fraction 683/2820, seems 
to be the length of the modern tropical year, mistaken 
for the duration of the vernal-equinox year. 
As a result,  the scheme runs into trouble many times during a 
period of 2025 years, for example 21 March A.D. -1, 21 March A.D. 1600, 
and 21 March A.D. 2025.   \\

The 2820-year cycle may be at the origin of the extremely precise 
duration of the year (365.24219858156 days) attributed to Khayy\^am 
(see above, Section 4). In fact, the fraction 683/2820 implies a 
year of 365.2421985815603 days, which surprisingly resembles the 
attributed value, thus casting doubt on its authenticity. This spurious value 
may stem from the mistake that Kh\^ayyam was the author of the 2820-year 
cycle. However, there is no historical record relating that cycle 
to Khayy\^am (Sayy\^ad 2000). The confusion is probably due 
to the fact that Behruz (1952) proposed his scheme in the form of 
tables which he called  "Khayy\^amic tables", leaving  
so the door open to  misinterpretation. \\

The conception of such a  cycle seems therefore to result from a 
confusion between 
the vernal-equinox and tropical years, as explained above, and 
moreover does not take into account the secular variation of the year length. 
Furthermore, this complex system does not make the calendar more 
precise, since the presently followed 33-year scheme, although much simpler, 
is quite efficient.  \\

\vspace*{1cm}
\parindent=0cm

{\bf 8. Correspondence with other calendars}  \\

\parindent=1cm 

There are several reliable software codes for converting the Iranian 
calendar dates and week days into other calendar systems and vice-versa. 
We can mention the {\it ``Khayam'' Program}: 

\begin{center}

http://payvand.com/calendar  \\

\end{center}

\parindent=0cm
set up by Hossein B\^agher-Z\^adeh for the correspondence between 
the Iranian and Gregorian calendars. 
Currently the leap years 
in the 33-year cycles  are those years that after
dividing by 33 leave a remainder of 1, 5, 9, 13, 17, 22, 26 and
30. For example, the year A.P. 1375 that begun on March 20, 1996 has the
remainder of 22 and thus is the leap year. These rules are implemented
in the ``Khayam program''. In a recent paper 
Borkowski (1996/1997) argues that the algorithm employed 
in that program is valid for the years A.D. 1799 to 2256
(A.P. 1178 to 1634). Moreover he presents a concise code which 
reconstructs the pattern of leap years over a time span of about 3000
years. \\

\parindent=1cm

Another interesting tool 
is the {\it Calendrica 2.0} software package, based on the algorithms in 
{\it Calendrical Calculations: The millennium Edition}, by Edward M. 
Reingold \& Nachum Dershowitz. The online version:  \\

\begin{center}
http://emr.cs.iit.edu/home/reingold/calendar-book/Calendrica.html  \\
\end{center}

\parindent=0cm

allows conversion not only between the Iranian and Gregorian 
systems, but also among several other calendars, mainly: Armenian, 
Chinese, French revolutionary (Note 16), Hebrew, Hindu, Islamic, Mayan, 
and the mistaken  arithmetic version of the Iranian calendar programmed 
with the 2820-year cycle of intercalations. \\

\vspace*{1cm}
\parindent=0cm

{\bf 9. Concluding remarks} \\

\parindent=1cm

In this paper, we presented a detailed description of the Iranian calendar 
emphasizing  the underlying astronomical phenomena, without 
ignoring the related historical topics. A prominent feature of the 
Iranian year is its starting point, Nowruz, which coincides 
with an astronomical event: the Sun's arrival at the vernal equinox. 
Nowruz, which symbolizes the rebirth of the nature and 
the triumph of life over the oppressive cold and darkness of the winter, 
is celebrated by a large number of peoples in the Western and Central 
Asia. \\

\parindent=1cm

On the basis of the ephemeris provided by the French IMCCE for the period 
A.D. --4000 to +2500, and the equations obtained by Bretagnon \& Rocher (2001) 
and Meeus (2002), we discussed the various definitions of the year and showed 
that the Iranian calendar is based upon the vernal-equinox year and not 
on the ``tropical  year''. \\

In fact the modern concept of the ``tropical'' year, formulated by 
researchers in the field of celestial mechanics, represents a mean year 
which does not depend on a particular starting point on the ecliptic. 
Moreover, it leaves out short-term fluctuations in the duration of the year. \\

Therefore, the ``tropical'' year of 365.2422 days does not at all 
correspond to the interval between two successive passages of the Sun 
through the vernal equinox. Attributing the value of 365.2422 days to 
such time interval is a mistake, unfortunately widely propagated. \\

In contrast, the Iranian, or real vernal-equinox, year relies on the 
Keplerian, non-uniform motion of the Earth around the Sun as well as 
on the short-term perturbations. This is why successive observations 
of the Sun's passage through the vernal-equinox are necessary in order 
to determine the starting instant of the new year and the real 
duration of the previous one. \\

Neglecting the short-term fluctuations, the mean length of the 
Iranian, or vernal-equinox, year at the present epoch is 365.2424 
solar days. This time length represents the mean interval between two 
successive passages of the Sun through the vernal equinox. \\

 The difference between the ``tropical'' and vernal-equinox years though
 is not limited to their length since their long-term variations is
 also unalike. \\

The confusion between both concepts is not only scientifically wrong, 
it may entail other mistakes and  can also undermine the accuracy of the 
Iranian calendar. However, since the importance of this issue is often 
not appreciated even among calendar experts and astronomers, in order to 
avoid the mistake, we proposed using two distinctive terms, ``tropical year'' 
for the newly introduced concept and ``vernal-equinox year'' when solar 
calendars, Iranian or Gregorian, are concerned. \\

The length of the year, as laid down by the intercalation system adopted 
by the Khayy\^am reform, is 365.2424.. days. This estimate agrees well 
with the real vernal-equinox year at Khayy\^am's epoch, which 
had a mean value of 365.2423 days, and is even in better agreement with the 
current length of the vernal-equinox year, whose mean value is 365.2424 days.
  \\

We also analyzed the unparalleled intercalation system of the Iranian 
calendar, which is based on a 33-year cycle, and pointed out the 
astronomical foundation of this cycle. There are two types of leap  
years in the Iranian calendar, quinquennial and quadrennial. 
The cycle contains 8 leap years, the first one being quinquennial and 
the seven others quadrennial. \\

Finally, we explained why the recently proposed 2820-year cycle is 
fallacious, and argued that  most probably it is due to a confusion between 
the vernal-equinox and ``tropical'' years. This cycle is 
not only erroneous, but also  such a complex intercalation scheme seems 
useless for the Iranian calendar, since the 33-year cycle is almost 
perfect.  We argued also that this scheme is at the origin of the 
extremely precise length of the solar year attributed to Khayy\^am. \\

\vspace*{1.cm}
\parindent=0cm

{\it \underline {Acknowledgements}}. It is a pleasure to thank the
French {\it Institut de M\'ecanique C\'eleste et de Calcul des
Eph\'em\'erides } (IMCCE), Director Dr. William Thuillot, and
Dr. Patrick Rocher for providing the ephemeris which was essential for
this study and also for helpful discussions.  My field of research in
astrophysics being well outside that of calendars and time reckoning,
I learned a lot in the process of writing this paper and would like to
thank all the people who took part in this project.  I am particularly
indebted to the calendar expert Mr. Simon Cassidy, Emeryville 
Ca. U.S.A., for numerous e-mail exchanges and for a critical reading 
of the paper which contributed to improve its content.  I express my 
deep gratitude to him.  I benefited also from discussions and comments 
by several colleagues at Paris Observatory, in particular Dr. Thibaut 
Le Bertre, Dr. Jean-François Lestrade, and Dr. James Lequeux. 
 I would like to thank them.   
I am grateful also to Dr. Vassilis Charmandaris, 
Cornell University, USA, for comments. 
My thanks are also directed to Mr. Fr\'ed\'eric Meynadier, 
Paris Observatory/Paris VI University, for his valuable and efficient 
cooperation. 
I am also thankful to Mr. Jean Meeus,  
Brussels, and  Dr. Iraj Malakpour, Geophysics Institute, Tehran, 
for replying to my questions.  
I would like also to acknowledge Drs. Edward M. Reingold \& 
Nachum Dershowitz as well as Dr. Hossein B\^agher-Z\^adeh 
for using  their calendar conversion codes. 
I am also grateful to Dr.  Kazimierz M. Borkowski, Toru\'n 
Radio Astronomy Observatory, Nicolaus Copernicus University, Poland, 
who read the paper and made  several interesting comments.
I want also to thank Dr. Jafar \^Agh\^ay\^ani-Ch\^avoshi, Sharif University, 
Tehran, for providing several historical documents. 
 I am also grateful to Dr. M.-Sch. Adib-Solt\^ani, Tehran, for his remarks, 
in particular regarding etymology. 
Finally, I would like to dedicate this paper to the memory of the 
late scholars  Zabih Behruz and Ahmad Birashk, who initiated the 
modern research on the Iranian calendar.  \\

\vspace*{1cm}

{\bf Notes}\\

{\bf 1.}
Etymology of Nowruz. Modern Persian {\it now} ``new'', 
Middle Persian {\it n\^ok}, Avestan {\it nava}, 
Sanskrit {\it nava}, akin to Greek {\it neos}, Latin {\it novus}, 
all from the Indo-Eurpean root {\it *newo}.  
The English {\it new, novel}, German {\it neu}, and French {\it 
nouveau, neuf} belong to this group.
The second component, 
Modern Persian {\it ruz} ``day'', Middle Persian {\it r\^oc},  
Old Persian {\it raucah}, Avestan {\it raocah} ``light, luminous; 
daylight'', Sanskrit {\it roka} ``brightness, light'', cognate with Greek 
{\it leukos} ``white, clear'', Latin {\it lux} ``light'' (also 
{\it lumen, luna}), Indo-Eurpoean root {\it *leuk} ``light, brightness''. 
The Persian words {\it rowshan} ``bright, clear'', {\it foruq} 
``light'', and {\it afruxtan} ``to light, kindle'' also belong to 
this family, as well as the English {\it light}, German {\it Licht}, 
and French {\it lumi\`ere}. \\

{\bf 2.}
There is no ambiguity between the vernal and autumnal equinoxes, even 
if the passage of the Sun through the autumnal equinox is the beginning 
of spring in the southern hemisphere! The vernal equinox is the point 
with coordinates 0,0 degree in the ecliptic system, whereas the autumnal 
equinox is defined by the point 180, 0  degrees. \\

{\bf 3.} 
Placing the beginning of the day at midnight is an ancient Iranian 
practice, as attested by several sources. For example,  
the famous scientist Abu-Rayh\^an Biruni (A.D. 973-1048) deals with 
this matter in his well-known book {\it Athar al-Baqia} (ancient history 
and geography),  written around A.D. 1000 (see below, references, 
for more details).  Moreover, he underlines  that the
{\it Shahriy\^ar  zij}, the astronomical calculations and tables 
established during the reign of the Sasanid emperor Khosrow I 
Anushirav\^an around A.D. 555, was based on the adoption of the  
midnight as  the beginning of the day. 
Note that taking up the midnight  for the beginning of 
the day by astronomers is relatively recent in the Western world.  
In fact until 1925 astronomers started and ended their days at noon, 
so the day in Greenwich Mean Time (GMT) originally started and ended at mean 
solar noon in Greenwich, while Greenwich Civil Time started at midnight. 
Nevertheless, the Julian days, used by astronomers, begin at noon (GMT). 
\\

{\bf 4.} According to various sources, up to eight astronomers  
participated in the reform project. 
Apart from Omar Khayy\^am, the other recorded names are: 
 Abu-H\^atam Mozaffar Esfaz\^ari, 
Abd-ol-Rahm\^an Kh\^azeni, Meymun ebn-e Najib V\^aseti, 
and Abol-Abbas Lukari. Some historians have brought into question the 
participation of  Kh\^azeni, a slave-boy of Byzantine 
origin, owned by Ali ebn-e  Kh\^azen Marwazi, treasurer 
and chancellor of the court at Marv (now Mary, Turkmenistan), who gave 
the young man 
the best possible education in mathematical and philosophical 
disciplines, so that he became a renowned mathematician/astronomer and 
``physicist''. Kh\^azeni established a {\it zij} for the ruler Sanjar, and 
invented a balance for measuring specific gravities which was 
as precise as those obtained up to the Eighteenth or beginning of the 
Nineteenth century. He passed away in A.D. 1115 or 1130, which is not 
incompatible with his taking part 
in the calendar reform. With the 
first date, he would have been 36 years old at the time of 
the reform, and 51 years with the second date. However, the intercalation 
system he proposes in the {\it Sanjari zij} does not fully agree with 
Khayy\^am's scheme (see also Note 15), and this may be interpreted as his 
non-adherence to the calendar reform. \\

{\bf 5.}  The current lengths of the astronomical seasons, around the 
year 2000, are about: spring 92.76 days, summer 93.65 days, autumn 
89.84 days, and winter 88.99 days. The season are unequal because the 
Earth's orbit is slightly elliptical and the Sun is not exactly at the 
center of the orbit. As formulated by Kepler's second law, the Earth 
moves faster when it is close to the Sun than when it is farther away, 
so the seasons that occur when the Earth is close to the Sun pass more 
quickly. The Earth reaches its perihelion, the point in its orbit  
closest to the Sun, in early January and is at aphelion, farthest away 
from the Sun, in early July. Hence the summer is longer than the 
winter in the Northern Hemisphere. In the Southern Hemisphere, the 
winter is longer than the summer. \\

The present situation is not, however, eternal and the duration of the 
seasons change through time due to variations in the orbital 
parameters of the Earth. The precession, or the motion of the 
orientation of the Earth's rotational axis, changes the positions of 
the solstices and equinoxes with respect to the perihelion. It takes 
about 21,000 years (climatic precession) for these points to make a 
complete circuit along the orbit, with respect to the perihelion. 
Similarly, variations in the shape, 
or the eccentricity, of the Earth's orbit (Note 12), which have a 
periodicity of about 100,000 years, modify the length of the 
seasons. And finally, changes in the obliquity, or tilt, of the 
Earth's axis cause important variations in the amount of 
solar radiation received by the Earth at high latitudes, bringing about 
considerable alterations in the seasonal climate.
The obliquity, which is currently about 23.5 degrees, varies between 
about 22.5 and 24.5 degrees over a period of approximately 41,000 years. \\

{\bf 6.} 
The Avestan, the language of the Avesta, scriptures of
Zoroastrianism,  belongs to the Iranian group of  the Indo-European 
family of languages. The oldest part  of the Avesta, called the 
{\it Gathas} (hymns, songs),  made up of poems attributed to Zarathushtra 
himself, are now commonly thought to date from around the end of the 2nd 
millennium BC and are thus contemporary with Vedic Sanskrit.\\

{\bf 7.} 
A brief etymological note. 
``Tropical'' from ``tropic'', from L.L. {\it tropicus} ``of or pertaining 
to the solstice'', from L. {\it tropicus} ``pertaining to a turn'', 
from Gk. {\it tropikos} ``of or pertaining to a turn
or change, or to the solstice'' (as a noun, ``the solstice''), from 
{\it tropos} ``turn, way, manner, style'', {\it tropein} ``to turn''. 
Indo-European root {\it *trep-} ``to turn''. Other terms of the same 
origin in English: troubadour, trover, contrive, retrieve, trophy, entropy. \\

The relation between the notions of ``tropic'' and ``solstice'' is due to the
observational fact that the Sun apparently ``turns back'' after reaching
its northernmost (or southernmost) point in the sky where it seems to 
stand still (solstice, from L. {\it solstitium}, 
from {\it sol} ``Sun'' + {\it -stit-, stes} ``standing'', akin to 
{\it stare}  ``to stand'', Greek {\it histanai} ``to cause to stand'', 
Sanskrit {\it sth\^a} ``to stand'', Avestan {\it st\^a} ``to stand'',  Persian 
{\it ist\^adan} ``to stand'', Indo-European root: {\it *st\^a} 
``to stand''). \\

The  term ``tropic'' is first attested around 1350-1400 in English with the 
meaning ``either of the two circles in the celestial sphere which describe 
the northernmost and southernmost points of the ecliptic''. Extended 1527
to the corresponding parallels of latitude on the 
terrestrial globe, one (tropic of Cancer) 23\deg\,27\min\, 
north, and the other (tropic of Capricorn) 23\deg\,27\min\, 
south of the equator, being boundaries of the Torrid Zone.
Meaning ``region between these parallels'' is
from 1837. ``Tropical'' first used 1520-1530, and ``tropical year'' is 
first attested 1585-1595. \\ 

For an interesting account of the history of the tropical year, 
see Meeus \& Savoie (1992). \\

{\bf 8.}
For example, the Royal Greenwich Observatory/ National Maritime Museum website:

\begin{center} 

{\footnotesize
http://www.nmm.ac.uk/site/request/setTemplate:singlecontent/contentTypeA/conWebDoc/contentId/349 }
\end{center}

 states that:\\

{\bf ``}The year is defined as being the interval between two successive
passages of the Sun through the vernal equinox. Of course, what is really
occurring is that the Earth is going around the Sun but it is easier to
understand what is happening by considering the apparent motion of the Sun
in the sky.\\

The vernal equinox is the instant when the Sun is above the Earth's
equator while going from the south to the north. It is the time which
astronomers take as the definition of the beginning of Spring. The year as
defined above is called the tropical year and it is the year length that
defines the repetition of the seasons. The length of the tropical year is
365.24219 days{\bf ''}. \\

We notice the widely spread confusion between the tropical and 
vernal-equinox years, leading to an erroneous value given for the 
mean interval between two successive passages of the Sun through the 
vernal equinox. The correct value is 365.2424 days. \\

Other examples of the same mistake: \\

{\bf ``}Tropical year: The year defined by two successive passages of the 
Sun through the vernal equinox: 365.242191 days{\bf ''} ({\it Dictionary 
of Science and Technology}, 1992, Academic Press, Inc.). \\

{\bf ``}Tropical year: The time interval between two successive passages of 
the Sun through the vernal equinox. Its length is 365.2422 mean 
solar days.{\bf ''} ({\it Encyclopedia of Astronomy and Astrophysics}, 2001, 
Institute of Physics Publishing). \\

{\bf 9.}  The Gregorian calendar is also comparable to the 
vernal-equinox year in the sense that the prime and stated aim of the 
Gregorian reform was to keep the instant of the spring equinox from 
drifting away from the date of 21 March. Since the first Council of 
Nicaea in A.D. 325, 21 March had been adopted as the ecclesiastical 
date of the spring equinox for celebrating Easter: the first Sunday 
following the first ecclesiastical full moon that occurs on or after 
the day of the vernal equinox. The discussion of the accuracy with 
which the Gregorian calendar manages to keep the vernal equinox on 21 
March is out of the scope of this note.\\

{\bf 10.}
The Delta-T parameter, which can only be deduced from observations, 
depends strongly on the rotation rate of the Earth, which is uncertain 
at millisecond level. The Delta-T expression used in the present IMCCE 
ephemeris is:

\begin{center}
{\it Delta-T = 102.3 + 123.5 T + 32.5 T$^{2}$ }   seconds \\
\end{center}

(Morrison \& Stephenson 1982), where $T$ is measured in centuries 
from the epoch J2000.0. Using a 
larger sample of observations of the historical solar and lunar 
eclipses, Stephenson \& Houlden (1986) derive a more accurate 
expression for the Delta-T parameter pertaining to the epochs 
before A.D. 948:

\begin{center}
{\it Delta-T = 2715.6 + 573.36 T + 46.5 T$^{2}$ }    seconds. \\ 
\end{center}
  
 Moreover, according to a recent review by Stephenson (2002), 
the terrestrial 
spin rate is not continually decreasing, but undergoes variations with 
a cycle of about 1500 years whereby we are now in a centuries long 
period of of no deceleration. Although the main causes of long-term 
changes in the length of the day are lunar and solar tides which 
produce a steady increase in the length of the day of approximately 
2.3 milli-seconds per century, there is a significant non-tidal 
component causing a secular decrease in the length of the day at a 
mean rate of 0.6 milli-seconds per century. Therefore, a simple 
parabolic analysis of Delta-T is not an ideal representation of the 
reality. These variations, if confirmed by future research, should not 
have an important effect on the averaged IMCCE data presented in this 
paper. As for the current (epoch +2000) year length measured in real 
solar days (Table 4, column 1), the truth probably lies somewhere 
between the solar and uniform-day values.  \\

{\bf 11.}
Meeus (2002) has called the vernal-equinox year the equinox-equinox 
year. This also is misleading since the vernal and autumnal years are 
not identical. \\

{\bf 12.} 
The shape of the Earth's orbit around the Sun, described by a  
parameter called {\it eccentricity}, changes periodically due to 
the gravitational pull of the planets. The eccentricity varies between 
0.070 (elongated ellipse) and 0.003 (almost circle) over some 
100,000 years. Currently the eccentricity is 0.017, but will become 
nearly circular in about 25,000 years. As a result, the difference 
between the tropical and vernal-equinox years will tend to zero. \\

{\bf 13.}
For more information about {\it Mithra} or {\it Mehr}, 
who has given his name to the 7th month of the Iranian calendar, and also 
its relation with Christmas and Marianne, the French symbol of the 
republic, see M. Heydari-Malayeri, {\it D'où vient le bonnet de Marianne?} 

\begin{center}
http://wwwusr.obspm.fr/$\sim$heydari/divers/marianne.html
\end{center}

{\bf 14.} 
For a discussion of  the name of the 4th Iranian month, 
see: M. Heydari-Malayeri, {\it Tishtar, the Iranian Sirius} (in preparation).
\\

{\bf 15.}
The fact that the 220-year cycle (with average year 365.2409 days)
may be partitioned into 29- and 25-year cycles (220 = 5 \x\,29 + 3 \x\,25),
would imply that Kh\^azeni had no use for the 33-year cycle
(365.2424 days/year). Therefore, if no other information is available 
from Kh\^azeni, one would have to conclude that either he has been 
miscopied in the extant texts or that he misunderstood or disbelieved 
Khayy\^am. \\

{\bf 16.}
According to several historical indications, the Iranian calendar has
been a source of inspiration for the creators of the French
revolutionary calendar, who preferred a non-religious time reckoning 
system, based
on natural seasons. As a matter of fact, during the age of enlightenment
European scholars set out to know the Eastern cultures
and civilizations in general. The first translation of the 
Zoroastrian sacred text Avesta in a European language was carried out 
by Abraham Anquetil-Duperron
in French in 1771. It is also highly probable  that the French
revolutionary thinkers were aware of the Iranian calendar through 
several published works. We can, for example, mention:
Freret, {\it De l'ancienne ann\'ee des Perses,} 1742, published in 
{\it l'Histoire de l'Acad\'emie Royale des inscriptions et belles 
lettres,} tome 16, Paris, 1751 and  Gilbert, {\it Nouvelles
observations sur l'ann\'ee des anciens Perses,} in {\it l'Histoire de
l'Acad\'emie Royale des inscriptions}, tome 31, Paris, 1788.
For more information, see the recent interesting paper by 
Shaf\^a (2003) who highlights the Persian elements in the French
revolutionary calendar. \\

%\vspace*{1.cm}

\newpage

{\bf References} \\

\begin{description}

\item
Behruz, Zabih, A.P. 1331 (A.D. 1952), {\it Taqvim va T\^arix dar 
Ir\^an} (Time reckoning \& calendar in Iran), Tehran, Iran-kudeh, 
No 15 (in Persian) 

\item
Birashk, Ahmad, 1993, {\it A Comparative Calendar of Iranian, 
Muslim Lunar, and Christian Eras for 3000 Years}, Mazda  Publishers,
Bibliotheca Persica, Costa Mesa, California and New York 

\item
Biruni, Abu-Rayh\^an , {\it c.} 1000, {\it Athar al-Baqia 'an al-Qurun 
al-Khalia},  translated into English under the title {\it The Chronology 
of Ancient Nations} by Edward C. Sachau, London 1879 (William H. Allen)

\item
de Bloy, Fran\cc ois, 1996, {\it The Persian Calendar}, J. Brit. Inst. 
   Persian Studies {\bf  34},  39

\item
Borkowski, Kazimierz M., 1996/1997, 
{\it The Persian calendar for 3000 years}, Earth, Moon and Planets 
{\bf No 3},223:

http://www.astro.uni.torun.pl/$\sim$kb/Papers/EMP/PersianC-EMP.htm

\item
Bretagnon, Pierre, Rocher, Patrick, 2001, {\it Du Tmps universel au Temps 
   coordonn\'ee barycentrique} Revue du Palais de la D\'ecouverte
   {\bf 285}, f\'evrier, p. 39

\item
Cassidy, Simon, 1996, {\it Error in Statement of Tropical Year}

http://www.angelfire.com/dc2/calendrics/index.html\#top

\item
  {\it Encyclopedia Iranica}, ed. Ehsan Yarshater, Columbia University, 
   Eisenbrauns, Inc., Winona Lake, Indiana 

\item
   Malakpour, Iraj, 2004, private communication.

\item
Meeus, Jean, 2002, {\it More Mathematical Astronomy Morsels}, 
William-Bell Inc., Richmond, Virginia, Ch. 63 

\item
Meeus, Jean, Savoie, Denis, 1992, {\it The history of the tropical year}, 
   J. Br. Astron. Assoc. {\bf 102}, 1 

\item
O'Connor, John J.,  Robertson,  Edmund F., 1999,  
The MacTutor History of Mathematics archive, Omar Khayyam, 
 
http://www-history.mcs.st-andrews.ac.uk/Mathematicians/Khayyam.html

\item
Morrison, L.V., Stephenson, F.R., 1982, {\it Sun and Planetary System}  
{\bf 96}, 73, Reidel, Dordrecht 

\item
Sayy\^ad, Mohammad R., 1981 (A.P. 1360), {\it The sequence of quadriennial 
   and quinquennial intercalations in the Jal\^ali calendar} 
   (in Persian), Proceedings of the 12-th Mathematics Conference, 
   Isfahacirc;n University, p. 33

\item
Sayy\^ad, Mohammad R.,  2000, {\it Farhang}, Quarterly J. of Humanities \& 
   Cultural Studies, vol. {\bf 12}, No. 29-32, Issue topics: 
   {\it Commemoration of Khayy\^am}, p. 53 (in Persian)  

\item
Seidelmann,  P. K., Guinot, B., Doggett, L.E., 1992, in 
   {\it Explanatory Supplement to the Astronomical 
   Almanac}, University Science Books, Mill Valley, Californian, 
   ed. P. K. Seidelmann, p. 80

\item
Shaf\^a, Shoj\^aedin (2003), in Persian Heritage, No {\bf 30} (t\^abestan-e 
A.P. 1382), p. 39 (in Persian)  

\item
Simon, J.L., Bretagnon, P., Chapront, J., Chapront-Touz\'e, M., 
   Francou, G., Laskar, J., 1994, {\it Astron. \& Astrophy.} {\bf 282}, 663 

\item 
Stephenson, F. Richard, 2003, Historical eclipses and Earth's rotation,
{\it Astronomy \& Geophysics} {\bf 44 (2)}, 2.22 

\item
Stephenson, F.R., Houlden, M.A., 1986, {\it Atlas of Historical 
   Eclipse Maps}, page x, Cambridge University Press
 
\item
Taqiz\^adeh, S. H., 1938, {\it Old Iranian Calendars}, Royal 
   Asiatic Society

   http://www.avesta.org/taqizad.htm

\item
Youschkevitch, A.P.,  Rosenfeld, B.A., 1973, ``al-Khayyami'', 
   {\it Dictionary of Scientific Biography} VIII, p. 323

\end{description}

\vspace*{1cm}

%Version 22/09/04 \\

\vspace*{1cm}

{\it Copyright 2004 by M. Heydari-Malayeri}  \\

\end{document}